# Modelling the Vibration-Rotation Energy Levels of $D_2^{18}O$ molecule with Effective Hamiltonian Method

I.A. Vasilenko, O.V. Naumenko, E.R. Polovtseva, A.D. Bykov

V. E. Zuev Institute of Atmospheric Optics SB RAS, Tomsk

Using the effective rotational Hamiltonian method, we have conducted an analysis of the $D_2^{18}O$ ground and the first excited vibration state rotational energy levels. The analysis was based on the effective Hamiltonians represented in several forms: the Watson Hamiltonian, the Hamiltonian expressed in terms of Padé-Borel approximants, and the Hamiltonian in terms of generating function expansions. The rotational and centrifugal constants have been determined from the fitting, which describe the rotational energy levels with an accuracy close to that of the experimental data. The predictive performance of the model with respect to highly excited rotational states has been evaluated against the global variation calculations. The radii of convergence of the effective rotation Hamiltonian series have been determined.

## 1. Introduction

A range of problems involving radiative processes in the Earth's atmosphere, astrophysical phenomena, development of isotope separation methods, synthesis of ultra-pure materials, and other problems of molecular physics require to use high-precision data on the absorption spectra of water vapour and its isotopic modifications. Such data can be obtained by modelling the vibration-rotation (VR) spectra of the water molecule, a typical asymmetric top, which is a challenging task to do, especially considering transitions to highly excited vibrational and rotational states.

Recently, a considerable progress in the precision of variational calculations of vibration-rotation energy levels of the water molecule and its isotopologues has been achieved [1-3]. The variational calculation, based on high-precision function of intramolecular potential energy determined by the least square fitting to the energy levels recoverable from spectra (e.g., see [2]), has the root-mean-square (RMS) accuracy of 0.02- 0.03 cm$^{-1}$ between the experimental values and the VR levels included in the fitting procedure. On the other hand, the variational calculations for the levels not included in the fitting give a much worse accuracy.

In parallel with the development of variational methods, the modelling of the VR energy levels based on the effective Hamiltonian approach has also been improving. The effective

Hamiltonian method is attractive. as it significantly reduces the problem's dimension and requires relatively small computations. Additionally, it is highly accurate for calculating rotational levels. However, this approach has a significant problem: the effective Hamiltonians are represented as series, and these series diverge for high rotational states. Application of various techniques developed for summation of divergent series appearing in the effective VR Hamiltonian resolves this issue, enabling the experimental data to be reproduced with a better accuracy, and the energy levels and the centres of the weak lines not observed experimentally to be predicted more reliably [4-5].

We present here the results of modelling the rotational spectrum of the ground and the first excited vibration states of the $D_2^{18}O$ molecule. The theoretical model uses different representations of the effective rotational Hamiltonians, namely, the Watson Hamiltonian, the Hamiltonian constructed from one-dimensional Padé-Borel approximation [4], and the Hamiltonian expressed in terms of the generating functions of centrifugal distortion [5]. For each of these Hamiltonians, their parameters have been fitted to the experimental energy levels determined from the spectrum. The results of the inverse problem solution have been cross-compared and the prediction accuracy for the rotational levels not included in the fit has been evaluated. Also, the divergent series summation methods enable one to determine the radii of convergence of the effective Hamiltonian series in terms of the rotational quantum numbers.

## 2. Experimental energy levels

In the present work, for the initial data we used the $D_2^{18}O$ energy levels obtained from analysing the high-resolution spectra in the range 969–2148 cm$^{-1}$ of deuterium-substituted water vapour enriched with the $^{18}O$ isotope. The spectra were recorded at the University of Science and Technology of China, Hefei, using a Fourier spectrometer with a resolution of about 0.001 cm$^{-1}$ equipped with a multi-pass gas cell [6-7]. The line positions for isolated lines of medium intensity were measured with an accuracy better than $4 \times 10^{-4}$ cm$^{-1}$. The energy levels ($J \leq 23$ and $Ka \leq 13$) were determined using the MARWEL method, in which the ground and exited levels are solutions to a system of linear equations [8].

## 3. Effective rotation Hamiltonian.

The Watson effective rotation Hamiltonian has the form [9]

$$H = H_D + \{J_{xy}^2, H_{ND}\}, \tag{1}$$

where the «diagonal» $H_D$ and «non-diagonal» $H_{ND}$ parts are represented as power series in angular momentum operators

$$H_D = E_V + (A^{[V]} - \frac{B^{[V]} + C^{[V]}}{2})J_z^2 + \frac{B^{[V]} + C^{[V]}}{2}J^2 - \\ - \Delta_k^{[V]}J_z^4 - \Delta_{jk}^{[V]}J_z^2 J^2 - \Delta_j^{[V]}J^4 + + H_k^{[V]}J_z^6 + ... \quad (2)$$

$$H_{ND} = \frac{B^{[V]} - C^{[V]}}{4} - \delta_k^{[V]}J_z^2 - \delta_j^{[V]}J^2 + h_k^{[V]}J_z^4 + ... \quad (3)$$

Conventional notations are used throughout this text. The effective rotation Hamiltonian parameters (the vibration energy $E_V$, the rotational constants $A^{[V]}, B^{[V]}, C^{[V]}$, the centrifugal constants $\Delta_k^{[V]}, \Delta_{jk}^{[V]}, \Delta_j^{[V]}, \delta_k^{[V]}, \delta_j^{[V]}, ...$ and others) are functions of the vibrational quantum number denoted by the superscript index *V*. For brevity, we shall call the method of calculations that uses expansions (2) and (3) the Watson model.

Since the series (2) and (3) may diverge, it is necessary to select an appropriate summation scheme. Various summation methods for the perturbation theory series have been previously proposed (see, for example, [11-22]). It has been demonstrated that these methods significantly improve calculations of the highly-excited vibration-rotation states. In the present work, the rotational energy levels of the ground (000) and the first excited (010) vibrational states of $D_2^{18}O$ molecule were modelled using both the conventional representation (1)-(3) of rotation Hamiltonian, and two additional Hamiltonian forms: the one obtained by the Padé-Borel summation and the other expressed in terms of the generating functions.

The Padé-Borel model uses the idea of a one-dimensional approximation of the effective Hamiltonian [4]. Under this model, the Hamiltonian is treated as a power series in a certain formal parameter λ. So that each matrix element of the Hamiltonian is a Borel-summed power series in λ, and the analytical continuation of the Borel image is found in the form a Padé sum. The formal parameter $\lambda$ is set to 1 in the final expression. The matrix elements of the effective Hamiltonian (1), when using one-dimensional Padé-Borel approximants

$$PB^{[1,1]}(\lambda) = \int_0^\infty e^{-t} P^{[1,1]}(\lambda t) dt \quad , \quad (4)$$

are expressed as follows:

$$\langle jk|H_D|jk\rangle = E_v + \int_0^\infty \frac{c_0 c_1 + (c_1^2 - c_0 c_2)t}{c_1 - c_2 t} e^{-t} dt$$

$$\langle jk|H_{ND}|jk\rangle = \int_0^\infty \frac{b_0 b_1 + (b_1^2 - b_0 b_2)t}{b_1 - b_2 t} e^{-t} dt \quad (5)$$

where

$$c_0 = \left[A^{[v]} - \frac{B^{[v]} + C^{[v]}}{2}\right]k^2 + \frac{B^{[v]} + C^{[v]}}{2} j(j+1)$$
$$c_1 = -\Delta_k^{[v]} k^4 - \Delta_{jk}^{[v]} k^2 j(j+1) - \Delta_j^{[v]} j^2(j+1)^2$$
$$2c_2 = H_k^{[v]} k^6 + H_{kj}^{[v]} k^4 j(j+1) + H_{jk}^{[v]} k^4 j^2(j+1)^2 + H_j^{[v]} j^3(j+1)^3 + L_k^{[v]} k^8 + ...$$

(6)

and

$$b_0 = \frac{B^{[v]} - C^{[v]}}{4}$$
$$b_1 = -\delta_k^{[v]}\left[k^2 + (k \pm 2)^2\right] - \delta_j^{[v]} j(j+1)$$
$$2b_2 = h_k^{[v]}\left[k^4 + (k \pm 2)^4\right] + h_{jk}^{[v]}\left[k^2 + (k \pm 2)^2\right]j(j+1) + h_j^{[v]} j^2(j+1)^2 + ...$$

(7)

In the left-hand side of (4), $P^{[1,1]}(z)$ signifies a Padé approximant of the order [1/1]. The integrals in (4) can be calculated exactly

$$\langle jk|H_D|jk\rangle = E_v + (c_0 c_2 - c_1^2)/c_2 + c_1 Ei(c_1/c_2)(c_1/c_2)^2 \exp(-c_1/c_2)$$

(8)

where $Ei(-x) = -\int_x^\infty e^{-t} t^{-1} dt$ is the integral exponential function. The matrix elements $\langle jk|H_{ND}|jk \pm 2\rangle$ are calculated in a similar way by replacing $c_n$ with $b_n$. Here $j$ and $k$ are respectively quantum numbers of total angular momentum and its projection onto the axis of the least moment of inertia.

The idea of generating functions, proposed in [5, 13, 21] is as follows. The effective rotation Hamiltonian (1) of an isolated vibrational state is a function of commuting operators $J_z, J^2$. If there exist functions such that their expansions in series coincide with the effective rotation Hamiltonian, they are called generating functions. Apparently, the generating functions are sums of the series (2) and (3) and can be obtained by applying a suitable series summation procedure. In the generating function model, the diagonal and non-diagonal parts are represented in the form:

$$H_D = \sum_m g_m(J) G_D(J^2, J_z^2)^m$$
$$H_{ND} = \sum_m u_m(J)\{J_{xe}^2, G_{ND}(J^2, J_z^2)^m\}$$

(9)

Where $G_D(J^2, J_z^2)$ and $G_{ND}(J^2, J_z^2)$ are "elementary" generating functions (often called G-functions),

$$G_D(J^2, J_z^2) = \frac{2}{\alpha(J)} \left\{ \sqrt{1 + \alpha(J)J_z^2} - 1 \right\}$$
$$G_{ND}(J^2, J_z^2) = \frac{2}{\beta(J)} \left\{ \sqrt{1 + \beta(J)J_z^2} - 1 \right\} \quad (10)$$

The parameters $\alpha(J)$ and $\beta(J)$, $g_m(J)$ and $u_m(J)$ are also expressed as series

$$\alpha(J) = \sum_n \alpha_n J^{2n}, \quad \beta(J) = \sum_n \beta_n J^{2n}$$
$$g_m(J) = \sum_n g_{mn} J^{2n}, \quad u_m(J) = \sum_n u_{mn} J^{2n} \quad (11)$$

The expansion coefficients $\alpha_n, \beta_n, g_{mn}, u_{mn}$ can be calculated from the series (2) and (3), but in practice they are derived by fitting the experimental energy levels.

### 3. Determination of rotational and centrifugal constants.

In the present work, we have redefined the effective Hamiltonian parameters for the states (000) and (010) using the Watson, Padé-Borel, and generating function models. The rotational and centrifugal constants were determined by the nonlinear least square method, on the assumption that the original rotational levels were given with equal accuracy. For both vibrational levels, (000) and (010), a total of 168 energy levels with $J \leq 12$ and $K_a \leq 12$ were included in the fitting. Rotational and centrifugal constants derived in the fitting process are given in Tables 1 - 4.

Each theoretical model used for calculations demonstrated a satisfactory agreement with the experimental data, giving an accuracy comparable with the accuracy of the experimental levels; this fact is confirmed by the RMS error at the bottom lines of Tables 1 and 3. All obtained spectroscopic parameters are statistically significant; the numbers in brackets denote 1σ confidence interval in the units of last significant digits.

### 5. Analysis of results

**5.1. Rotational and centrifugal constants.** It should be noted that the effective Watson Hamiltonian makes it possible to achieve an accuracy of reproduction of the experimental energy levels, which is quite comparable with the Padé - Borel methods and generating functions. Thus for the ground vibrational state, the RMS errors for the Watson Hamiltonian and the generating function methods are almost equal, 1.55 10$^{-4}$ and 1.44 10$^{-4}$ cm$^{-1}$, respectively. While the RMS deviation for the Padé-Borel model is larger by a factor of about two, 2.52 10$^{-4}$ cm$^{-1}$. For the first excited vibrational level, the picture is similar- all three theoretical models give approximately the same results for the RMS deviations: 1.4 10$^{-4}$, 2.6 10$^{-4}$, and 4.0 10$^{-4}$ cm$^{-1}$. The rotational and centrifugal constants $\Delta$ are in good agreement throughout the models (see Tables 1 and 3). However, the higher order constants differ significantly. For example, the constants $P_k$ for the three models may differ by orders of magnitude and even have different signs: 1.5668(762)×10$^{-10}$ in the Watson model, 5.803(580) × 10$^{-12}$ in the Padé-Borel model, and -2.1114×10$^{-12}$ in the generating function model. Other centrifugal constants behave similarly. It turns out that high-order centrifugal constants are model-dependent and it is not possible to favour one model over the other. A more reliable conclusion can be made if one knows predictive properties of the models for higher energy states.

**5.2. Extrapolative properties of the effective Hamiltonians.** Let us consider calculations for higher energy levels not included in the fitting. The extrapolative properties of the rotation Hamiltonian parameters in the Watson, Padé-Borel, and generating function models were evaluated by comparing the calculated levels with the corresponding levels derived as a result of the "global" variational calculations with a high-precision potential energy function. The long practice of usage of variational calculations for water molecule has shown that these calculations have the best extrapolative properties. The results of comparison are shown in Fig. 1.

When extrapolating for large quantum numbers $K_a$ (maximum $K_a$ in the fitting procedure was 12 for J = 12), up to J = 20 and $K_a$ = 20, within the standard Watson model, the calculated VR levels deviated by up to 100 cm$^{-1}$ from high-precision variational calculations (Fig. 1a). Maximum deviation for the Padé-Borel model was less than 3 cm$^{-1}$ (Fig. 1c). The most accurate predictive calculations were obtained for the generating function model, which agreed with the variational calculations with an accuracy of 0.4 cm$^{-1}$ (Fig. 1e). It is noteworthy here to underline some points which are deemed important for analysis of the results.

As seen from Fig. 1a, for small values of the quantum number $K_a$ ~0, 1, predictive calculations with the Watson Hamiltonian give a satisfactory result – the deviation is hundredths of reciprocal centimetre. For larger quantum numbers of $K_a$~18-20, the energy levels obtained from the Watson Hamiltonian are by up to 92 cm$^{-1}$ lower that those obtained from the variational method. Meanwhile, for the levels with lower values of $K_a$ <17, the conventional Watson effective Hamiltonian, which is represented by divergent series, gives a fairly acceptable predictions of the energy levels. Maximum deviation from variational calculations is as small as 0.6 cm$^{-1}$ for 13≤ $J$ ≤15 and about 2 cm$^{-1}$ for $J$ =16, $K_a$=16.

The use of Padé-Borel approximants (4)-(9) significantly improves the extrapolative performance of the effective Hamiltonian. Maximum deviation (up to 2.58 cm$^{-1}$) from variational calculations is observed for the levels with $K_a \sim J$; while for the levels with $J \leq 18$, the error does not exceed 1 cm$^{-1}$ (Fig. 1c). It should be noted that the Padé-Borel model is based on the simplest first-order Padé approximation [4]. Application of higher order Padé approximants might improve the extrapolative properties of the effective Hamiltonian.

The most accurate results were obtained when using the generating function method (Fig. 1e). Maximum deviation from variational calculations is 0.36 cm$^{-1}$ for levels [20 15 6] and [20 15 5]. Note that dependence of the prediction error on the quantum number $K_a$ is not monotonic. It means that the method needs to be improved to account for details of the rotational energy for given quantum numbers. A similar picture is observed for rotational levels of the state (010), Figs. 1b, 1d, and 1f.

**5.3. Rotational energy singularities and radii of convergence.** It is of interest to determine singularities of the functions that are included in the effective rotation Hamiltonian. The functions $H_D$ and $H_{ND}$ can be treated as functions of the complex variables $J^2$ and $J_z^2$ so that one can make use of the known properties of analytical functions to calculate the energy levels. In analysing the analytical properties of rotational energies, we will use the parameters obtained for D$_2^{18}$O in the present work, as well as the data for H$_2^{16}$O referenced in [13]. The analysis was carried out using the standard methods discussed in [19]. The analytical properties of the effective rotational Hamiltonian series have been previously studied in [5, 21].

In the Padé-Borel model (4) - (5), the location of singular points is determined by the function $Ei(z)$. As known, this function has a logarithmic branch point $z=0$, as a consequence, the series $H_D$ and $H_{ND}$ are assumed to be divergent, having a radius of convergence given by the equation $c_1 = 0$ or $b_1 = 0$.

Since the equations (9)-(11), along with the coefficients $g_{mn}, u_{mn}$, reproduce the rotational energy spectrum with a high accuracy and possess, as shown above, satisfactory extrapolative properties, one may assume that they also give the singularities of $H_D$ and $H_{ND}$ on the complex plane. In such approximation, the study of analytical structure of rotational and centrifugal energy is greatly simplified, since it can be reduced to studying singularities of the approximants. Thus, for example, the radicand in (10) determines the branch points that in their turn determine the radii of convergence in $J_z$ of the series (2) and (3), asymptotic behaviour of higher orders and other properties of these series [19, 22]. It is easy to see that the branch points $J_z^{(c)}$ of the elementary generating functions (10) are purely imaginary ($J_z^{(c)} = \pm i/\sqrt{\alpha(J)}$ for the diagonal part and $J_z^{(c)} = \pm i/\sqrt{\beta(J)}$ for non-diagonal). Except for the quadric branch points, there are also the poles determined by the equations $\alpha(J)=0, \beta(J)=0$.

It should be underlined the following. According to (9), the generating function method requires to consider two series, $H_D$ and $H_{ND}$, and, hence two generating functions. It follows from the analysis of the rotational energy levels of the $H_2^{16}O$ vibrational states of type $(0v_20)$ that the two functions, whose parameters are determined as a result of fitting to the experimental level, are significantly different. They have different singular points and hence different radii of convergence. As an example, Fig. 2 shows the radii of convergences in $J_z$ which are defined as absolute values of the singular points of the generating functions (the ground state of $H_2^{16}O$). As seen from the figure, the radius of convergence of $G_D(J^2, J_z^2)$ is greater than that of $G_{ND}(J^2, J_z^2)$. Thus the divergence of the effective Hamiltonian for small values of J is determined by the non-diagonal part $H_{ND}$, namely, by the quadric branch points of $G_{ND}(J^2, J_z^2)$.

The expressions (9) can be regarded as function definitions in terms of expansions in series with the coefficients $g_m(J), u_m(J)$. These functions may have their own singularities which will affect the effective Hamiltonian properties. To determine these singularities, we used the second order Padé-Hermite approximants separately for $H_D$ and $H_{ND}$. The quadric branch points of these approximants are determined as the roots of a quadric equation (e. g., see [17-20, 22]). As mentioned above, it is assumed that the singularities of the approximants coincide with the singularities of $H_D$ and $H_{ND}$.

The radii of convergence shown in Fig. 2 depend weakly on the quantum number $J$ and are stable, which confirms correctness of their determination. The radii of convergence relating to the singular points of $G_D(J^2, J_z^2)$ and $G_{ND}(J^2, J_z^2)$ turn out to be smaller than those of the $H_D$ and $H_{ND}$ series. Thus the dominant singularities of the rotational energy have been accounted for in the generating functions, which explains good extrapolative properties of the effective Hamiltonian expressed in the form (9).

Note that for $H_2^{16}O$ the radius of convergence of the diagonal part of the effective Hamiltonian have been earlier found to be equal 10 using a simplified model of bending vibration-rotation interaction [5], which is somewhat greater than the radius determined in the present work. The Hamiltonian that includes the Padé-Borel approximants models the singular point as logarithmic branch points which implies a zero radius of convergence of the series (2) and (3). This results in a poorer extrapolation properties as compared with the generating function method.

A similar analysis of rotational energy singularities has been conducted for the ground and the first excited vibrational states of the $D_2^{18}O$ molecule based on our data (Tables 2 and 4) and the data in Ref. [7]. The obtained radii of convergence for the angular momentum values $J = 0,…, 20$ are shown in Fig. 3. It is noteworthy that, as in the case of the $H_2^{16}O$ ground state, the divergence is defined by the singularities of the elementary generating function (circles in Fig. 3). Similar results are true for the vibrational state (010).

It should be noted that the «elementary» generating functions (10) coincide with the quadric Padé-Hermite approximants of the order [0, 0, 1]. In [5], an elementary generating function of the form $G_D(J^2, J_z^2)$ for the diagonal part of the effective Hamiltonian was derived as a solution to a model problem that approximately describes bending - rotation interaction in the molecule of $H_2X$ ($C_{2V}$) - type. Defining a similar simple theoretical model for non-diagonal part of the Hamiltonian seems a challenging task. An acceptable solution can be to define $G_{ND}(J^2, J_z^2)$ as one of the Padé-Hermite approximants.

## 6. Conclusion

In the present work, we have presented the results of modelling the rotational energy spectrum of the ground and the first excited vibrational states of a heavy isotope modification of $D_2^{18}O$. By the least square fitting to the experimental energy levels, we have recovered the rotational and centrifugal constants of the effective Hamiltonians represented in different forms. The calculated energy levels agree well with the experimental values for all three theoretical models.

Predictive abilities of the different Hamiltonians for long extrapolations in the quantum number $K_a$ have been evaluated by comparing with variational calculations for the $D_2^{18}O$ vibrational states (000) and (010). As expected, the extrapolation with the Watson Hamiltonian, while having a relatively high fitting accuracy, has proved to give the worst results because of the divergent series that represent the Hamiltonian. On the contrary, the methods based on summation of divergent series give a much better agreement with the variational calculations. Of all the considered methods, the generating function approach has proved more preferable for a long-range extrapolation.

We have determined the branch point of the rotational energy on the complex plane of variable $J_z$. (corresponding to the quantum number $k$). These points determine the radii of convergence of the effective Hamiltonian series (2) and (3). It has been found that for small values of the quantum number $J$ the series divergence is determined by the non-diagonal part of the effective Hamiltonian $H_D$.

The work was supported by RFBR under Grant No. 18-02-00462.

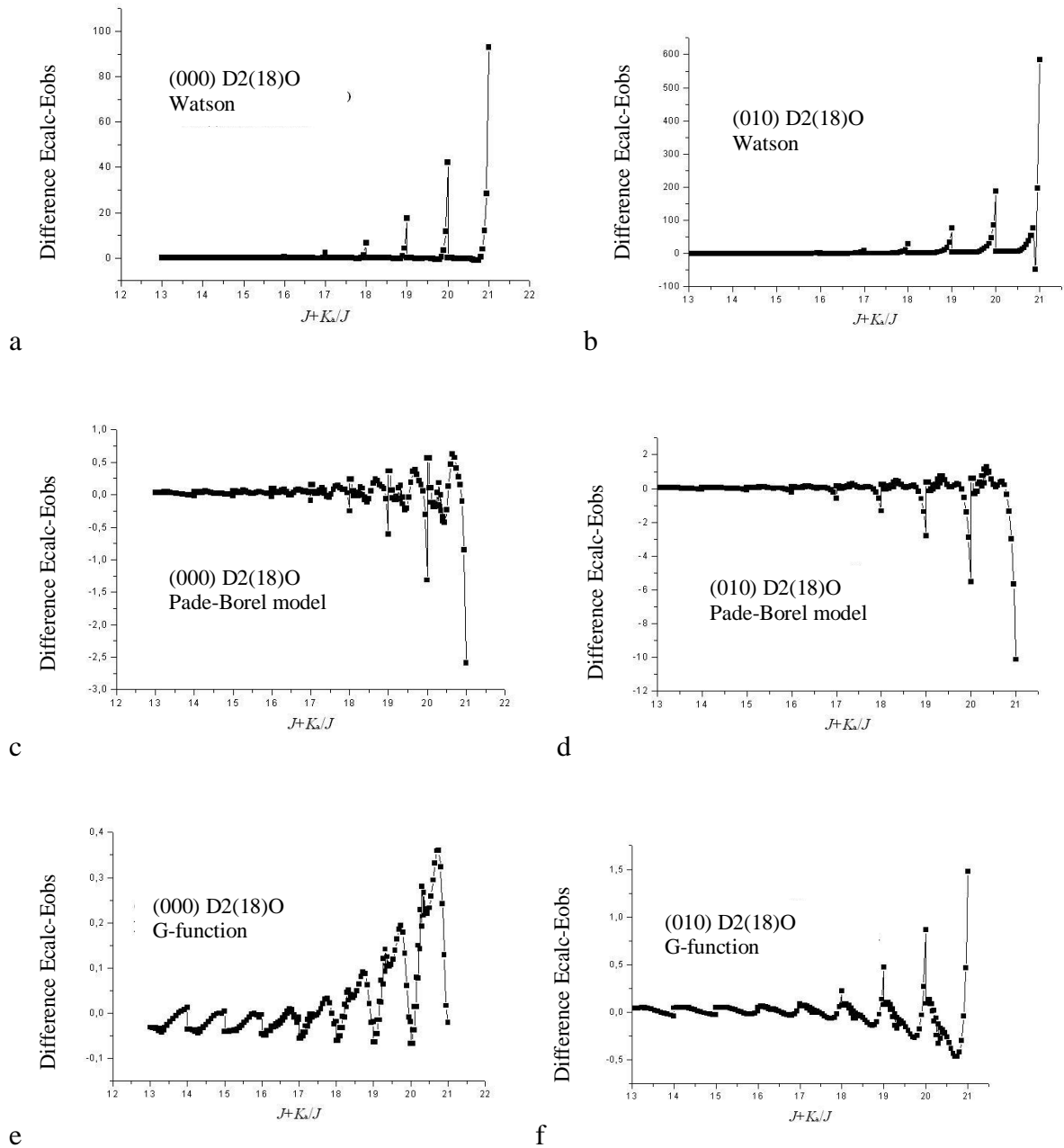

Figure 1. The energy levels $13 \leq J \leq 20$ of the ground and the first excited vibrational states of the $D_2^{18}O$ molecule calculated using the effective Hamiltonians: (a) and (b) – the Watson Hamiltonian, (c) and (d) – Padé-Borel model, (e) and (f) - generating function model, in comparison with the levels $E_{var}$ obtained from variational calculations.

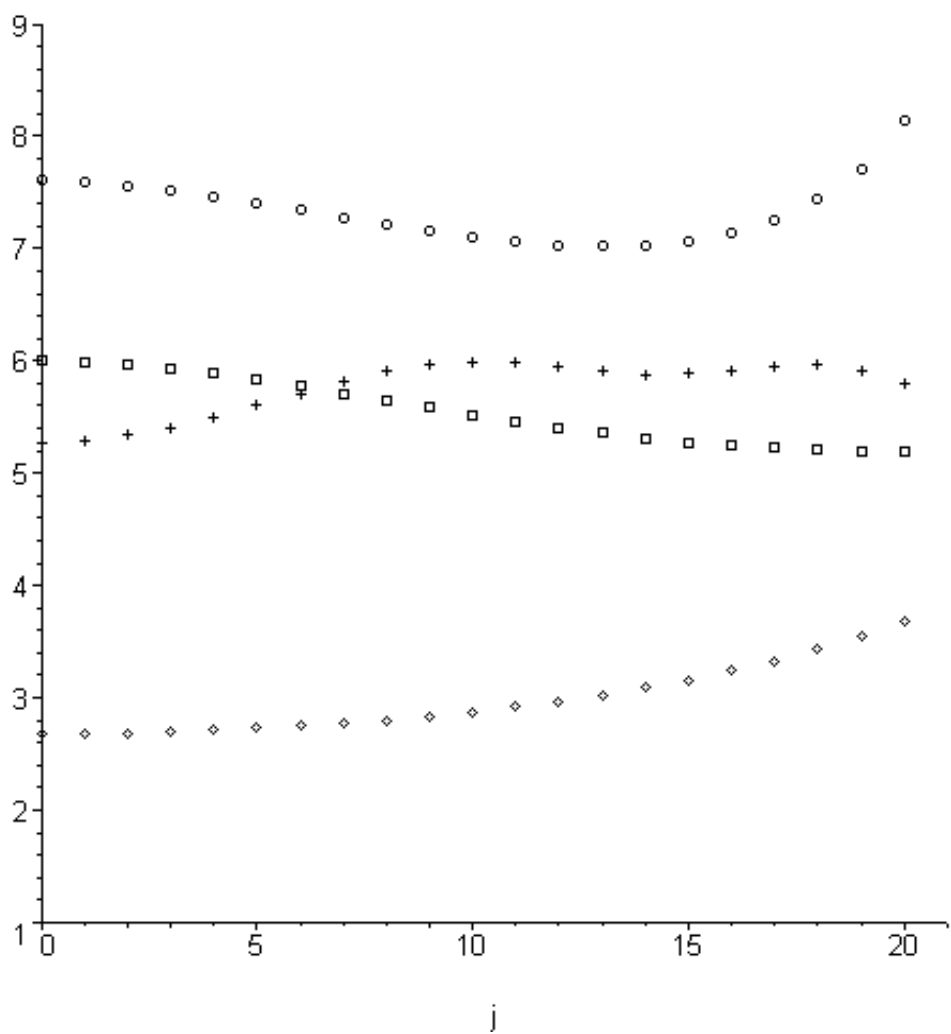

Figure 2. The radii of convergence in quantum number *k* for the effective Hamiltonian series (2)-(3) of the $H_2^{16}O$ ground vibrational state. The total angular momentum j is plotted along the abscissa axis. The radii of convergence are marked as circles for the $H_D$ series, crosses - for the $H_{ND}$ series, squares denote the radii of convergence determined for the elementary function $G_D(J^2, J_z^2)$, diamonds for $G_{ND}(J^2, J_z^2)$.

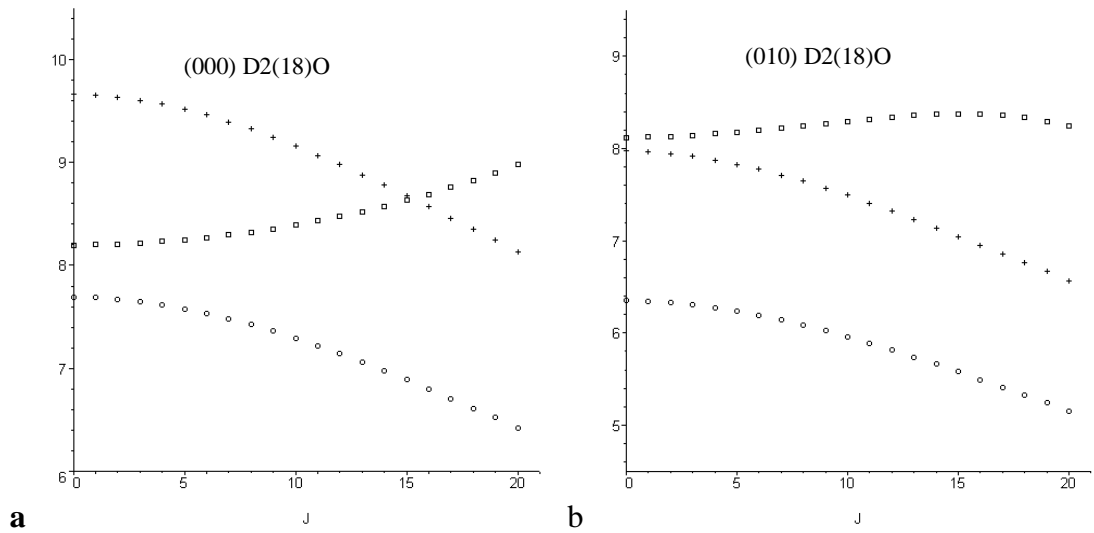

Figure 3. The effective Hamiltonian radii of convergence. a – the ground state, b – the state (010) of $D_2^{18}O$, circles for the radii determined by the elementary function $G$ (10); crosses for the radii of convergence of the $H_D$ series (9); crosses for the radii of convergence of the $H_{ND}$ series, (9).

Table 1. Rotational and centrifugal distortion constants of the $D_2^{18}O$ ground (000) state (cm$^{-1}$)

| Parameter | Watson Hamiltonian | Padé-Borel approximants | G-functions |
|---|---|---|---|
| A | 15.073474(142) | 15.073447(16) | 15.0734 |
| B | 7.2731814(602) | 7.2732261(18) | 7.2731 |
| C | 4.8100463(437) | 4.8100650(926) | 4.8100 |
| $\Delta_k$ | 8.85137(121) 10$^{-3}$ | 8.85473(130) 10$^{-3}$ | 8.849 10$^{-3}$ |
| $\Delta_{jk}$ | -1.492311(527) 10$^{-3}$ | -1.496982(811) 10$^{-3}$ | -1.4882 10$^{-3}$ |
| $\Delta_j$ | 3.0864217(417) 10$^{-4}$ | 3.10345(239) 10$^{-4}$ | 3.0869 10$^{-4}$ |
| $\delta_k$ | 3.22811(431) 10$^{-4}$ | 3.18955(735) 10$^{-4}$ | 3.2299 10$^{-4}$ |
| $\delta_j$ | 1.230764(417) 10$^{-4}$ | 1.235774(895) 10$^{-4}$ | 1.2305 10$^{-4}$ |
| $H_k$ | 1.72056(376) 10$^{-5}$ | 1.68177(453) 10$^{-5}$ | 1.6923 10$^{-5}$ |
| $H_{kj}$ | -2.4863(288) 10$^{-6}$ | -1.7987(322) 10$^{-6}$ | -2.2380 10$^{-6}$ |
| $H_{jk}$ | -2.4420(383) 10$^{-7}$ | -4.8762(125) 10$^{-7}$ | -2.0651 10$^{-7}$ |
| $H_j$ | 5.9656(284) 10$^{-8}$ | 8.228(264) 10$^{-8}$ | 5.9438 10$^{-8}$ |
| $h_k$ | 3.55384(194) 10$^{-6}$ | 2.1935(906) 10$^{-6}$ | 3.8021 10$^{-6}$ |
| $h_{jk}$ | -4.877(326) 10$^{-8}$ | | -8.1649 10$^{-8}$ |
| $h_j$ | 2.9614(183) 10$^{-8}$ | 3.4562(498) 10$^{-8}$ | 2.9586 10$^{-8}$ |
| $L_k$ | -5.0388(737) 10$^{-8}$ | -0.7811(315) 10$^{-8}$ | -4.8352 10$^{-8}$ |
| $L_{kkj}$ | 1.0411(810) 10$^{-8}$ | 0.6571(443) 10$^{-8}$ | 1.5925 10$^{-8}$ |
| $L_{kjk}$ | | -5.891(298) 10$^{-9}$ | -5.6333 10$^{-9}$ |
| $L_{jk}$ | | 1.3583(671) 10$^{-9}$ | |
| $L_j$ | | -6.370(857) 10$^{-11}$ | |
| $l_k$ | -0.15008(620) 10$^{-8}$ | | -3.2107 10$^{-8}$ |
| $l_{kj}$ | | 1.1982(662) 10$^{-8}$ | 1.7565 10$^{-9}$ |
| $l_{jk}$ | | -6.562(403) 10$^{-10}$ | |
| $P_k$ | 1.5668(762) 10$^{-10}$ | 5.803(580) 10$^{-12}$ | -2.1114 10$^{-12}$ |
| $P_{kkkj}$ | -5.339(868) 10$^{-11}$ | | -5.7098 10$^{11}$ |
| $p_k$ | 7.253(549) 10$^{-11}$ | 2.394(107) 10$^{-10}$ | 3.3892 10$^{-10}$ |
| $p_{kkj}$ | | -1.1898(654) 10$^{-10}$ | -4.8072 10$^{-11}$ |
| $Q_k$ | -3.444(195) 10$^{-13}$ | | 3.9429 10$^{-12}$ |
| $Q_{kkkkj}$ | 1.963(194) 10$^{-13}$ | | -9.1597 10$^{-13}$ |
| RMS | 1.55 10$^{-4}$ | 2.52 10$^{-4}$ | 1.44 10$^{-4}$ |

Table 2. Effective Hamiltonian parameters (9)-(11) of the generating function method for the $D_2^{18}O$ vibrational state (000).

| Parameter | Value | 1σ confidence interval |
|---|---|---|
| $\alpha_1$ | $0.168894616988 \times 10^{-01}$ | 0.36 10-03 |
| $\alpha_2$ | $0.174841093680 \times 10^{-4}$ | 0.45 10-06 |
| $g_{10}$ | 6.04162083479 | 0.22 10-05 |
| $g_{20}$ | $-0.308696427217 \times 10^{-3}$ | 0.46 10-07 |
| $g_{30}$ | $0.594386779578 \times 10^{-7}$ | 0.25 10-09 |
| $g_{01}$ | 9.03187795160 | 0.13 10-04 |
| $g_{11}$ | $0.148828700396 \times 10^{-2}$ | 0.45 10-06 |
| $g_{21}$ | $-0.206515312145 \times 10^{-6}$ | 0.34 10-08 |
| $g_{02}$ | $0.292868247064 \times 10^{-1}$ | 0.82 10-03 |
| $g_{12}$ | $0.435246474948 \times 10^{-4}$ | 0.12 10-05 |
| $g_{03}$ | $-0.578049325606 \times 10^{-4}$ | 0.16 10-05 |
| $g_{13}$ | $-0.803335069499 \times 10^{-7}$ | 0.26 10-08 |
| $g_{04}$ | $0.824882043296 \times 10^{-8}$ | 0.14 10-08 |
| $g_{05}$ | $0.863712166349 \times 10^{-10}$ | 0.65 10-11 |
| (B-C)/4 | 0.615785046140 | 0.21 10-05 |
| $u_{10}$ | $-0.123051295481 \times 10^{-3}$ | 0.38 10-07 |
| $u_{20}$ | $0.295860409121 \times 10^{-7}$ | 0.17 10-09 |
| $u_{01}$ | $-0.322993973930 \times 10^{-3}$ | 0.43 10-06 |
| $u_{11}$ | $-0.816495791008 \times 10^{-7}$ | 0.31 10-08 |
| $u_{02}$ | $0.243831330472 \times 10^{-5}$ | 0.20 10-07 |

*Note:* $\alpha_n$ are dimensionless, other parameters are in cm$^{-1}$.

Table 3. Rotational and centrifugal distortion constants of the $D_2^{18}O$ state (010) (cm$^{-1}$)

| Parameter | $J_z^n\ J^m$ | Watson | Padé-Borel | G - functions |
|---|---|---|---|---|
| $A$ | | 16.2560250(316) | 16.255990(174) | 16.2560 |
| $B$ | | 7.3370467(165) | 7.3370940(150) | 7.3370 |
| $C$ | | 4.7550856(126) | 4.755160(135) | 4.7551 |
| $\Delta_k$ | 4 0 | 1.336128(270)10-2 | 1.338708(159) 10$^{-2}$ | 1.33844 10$^{-2}$ |
| $\Delta_{jk}$ | 2 2 | -1.84116(151)10-3 | -1.862267(741) 10$^{-3}$ | -1.8554 10$^{-3}$ |
| $\Delta_j$ | 0 4 | 3.3087(164)10-4 | 3.35115(103) 10$^{-4}$ | 3.3455 10$^{-4}$ |
| $\delta_k$ | 2 0 | 7.06872(509) 10$^{-4}$ | 7.0737(177) 10$^{-4}$ | 7.0597 10$^{-4}$ |
| $\delta_j$ | 0 2 | 1.36377(114) 10$^{-4}$ | 1.3631(210) 10$^{-4}$ | 1.3606 10$^{-4}$ |
| $H_k$ | 6 0 | 3.8597(822)10-05 | 4.0057(453) 10$^{-5}$ | 3.9576 10$^{-5}$ |
| $H_{kj}$ | 4 2 | -5.3459(393)10-6 | -5.7918(239) 10$^{-6}$ | -5.4900 10$^{-6}$ |
| $H_{jk}$ | 2 4 | 3.943(284)10$^{-7}$ | 7.974(814) 10$^{-8}$ | 1.6004 10$^{-7}$ |
| $H_j$ | 0 6 | | 7.9638(510) 10-8 | 7.2197 10$^{-8}$ |
| $h_k$ | 4 0 | 8.0612(448) 10$^{-6}$ | 8.793(208) 10$^{-6}$ | 9.7296 10$^{-6}$ |
| $h_{jk}$ | 2 2 | | | -1.7315 10$^{-7}$ |
| $h_j$ | 0 2 | 3.6247(516) 10$^{-8}$ | 3.600(185) 10$^{-8}$ | 3.5851 10$^{-8}$ |
| $L_k$ | 8 0 | -1.2602(111)10$^{-7}$ | -2.6731(385) 10$^{-8}$ | -1.7138 10$^{-7}$ |
| $L_{kkj}$ | 6 2 | | 1.2874(430)10$^{-8}$ | 4.9492 10$^{-8}$ |
| $L_{kjk}$ | 4 4 | | -5.749(182) 10$^{-9}$ | -1.5195 10$^{-8}$ |
| $L_{jk}$ | 2 6 | -1.352(137) 10$^{-9}$ | | -1.2413 10$^{-10}$ |
| $L_j$ | 0 8 | 0.5471(152) 10$^{-10}$ | | |
| $l_k$ | 6 0 | | 7.414(199) 10$^{-8}$ | -1.2052 10$^{-7}$ |
| $l_{kj}$ | 4 2 | | -2.385(278) 10$^{-8}$ | 6.4769 10$^{-9}$ |
| $l_{jk}$ | 2 4 | | 1.416(301) 10$^{-9}$ | |
| $P_0$ | 10 0 | 4.37578(833)10-10 | 2.3768(712) 10$^{-11}$ | 1.6268 10$^{-10}$ |
| $P_2$ | 8 2 | -1.857(108) 10$^{-10}$ | | -3.4865 10$^{-10}$ |
| $P_4$ | 6 4 | 1.3619(631) 10$^{-10}$ | | 2.7229 10$^{-10}$ |
| $P_6$ | 4 6 | | | -4.5633 10$^{-13}$ |
| $P_{10}$ | 0 10 | -1.3821(701)-12 | | |
| $p_0$ | 8 0 | | 4.387(186) 10$^{-10}$ | 1.9913 10$^{-9}$ |
| $p_2$ | 6 2 | -1.1187(571) 10$^{-10}$ | -5.153(143) 10$^{-10}$ | -2.2920 10$^{-10}$ |
| $p_4$ | 4 4 | | 1.625(104) 10$^{-10}$ | 7.0941 10$^{-12}$ |
| $p_6$ | 2 6 | | -1.175(155) 10$^{-11}$ | |
| $p_8$ | 0 8 | | 1.129(278) 10$^{-13}$ | |
| $Q_0$ | 12 0 | -7.083(195) 10$^{-13}$ | | 2.5008 10$^{-11}$ |
| $Q_2$ | 10 2 | | | -3.0696 10$^{-12}$ |
| $Q_4$ | 8 4 | 4.706(384) 10$^{-13}$ | | -4.3083 10$^{-12}$ |
| $Q_6$ | 6 6 | -3.531(261) 10$^{-13}$ | | 2.3830 10$^{-13}$ |
| $Q_8$ | 4 8 | | | 9.5032 10$^{-16}$ |
| RMS | | 4.0 10$^{-4}$ | 2.6 10$^{-4}$ | 1.4 10$^{-4}$ |

Table 4. Effective Hamiltonian parameters (9)-(11) of the generating function method for the $D_2^{18}O$ vibrational state (010).

| Parameter | Value | 1σ confidence interval |
|---|---|---|
| $\alpha 1$ | 0.247757237070× 10-01 | 0.52×10-03 |
| $\alpha 2$ | 0.306215187342× 10-04 | 0.89 ×10-06 |
| E | 1170.15723863 | 0.58× 10-04 |
| g10 | 6.04612134232 | 0.41× 10-05 |
| g20 | -0.334556362640× 10-03 | 0.71× 10-07 |
| g30 | 0.721970347802× 10-07 | 0.34× 10-09 |
| g01 | 10.2099333913 | 0.20× 10-04 |
| g11 | 0.185544091183× 10-02 | 0.79× 10-06 |
| g21 | 0.160049412084× 10-06 | 0.82× 10-08 |
| g31 | -0.124138616656× 10-09 | 0.28× 10-10 |
| g02 | 0.498551701245× 10-01 | 0.13× 10-02 |
| g12 | 0.841633799374× 10-04 | 0.25× 10-05 |
| g03 | -0.126228146191× 10-03 | 0.36× 10-05 |
| g13 | -0.223443496277× 10-06 | 0.75× 10-08 |
| g04 | 0.505247685247× 10-07 | 0.28× 10-08 |
| g05 | 0.471475975167× 10-09 | 0.28× 10-10 |
| (B-C)/4 | 0.645476496844 | 0.22× 10-05 |
| u10 | -0.136061608565× 10-03 | 0.42× 10-07 |
| u20 | 0.358516295341× 10-07 | 0.19× 10-09 |
| u01 | -0.705972718367× 10-03 | 0.69× 10-06 |
| u11 | -0.173152650779× 10-06 | 0.37× 10-08 |
| u02 | 0.535686975056× 10-05 | 0.37× 10-07 |
| u22 | 0.576859094310× 10-11 | 0.84× 10-12 |
| u04 | 0.124944941797× 10-09 | 0.40× 10-11 |

*Note:* $\alpha_n$ are dimensionless, other parameters are in cm$^{-1}$.